\documentclass[twocolumn,english,prl]{revtex4-1}
\usepackage[T1]{fontenc}
\usepackage[latin9]{inputenc}
\usepackage[a4paper]{geometry}
\usepackage{bm}
\usepackage{amsmath}
\usepackage{amssymb}
\usepackage{stackrel}
\usepackage{graphicx}
\usepackage{babel}
\begin{document}

\title{Two impurities in a Bose-Einstein condensate:\\
 from Yukawa to Efimov attracted polarons}

\author{Pascal Naidon}

\affiliation{RIKEN Nishina Centre, RIKEN, Wak{\=o}, 351-0198 Japan}
\begin{abstract}
The well-known Yukawa and Efimov potentials are two different mediated
interaction potentials. The first one arises in quantum field theory
from the exchange of virtual particles. The second one is mediated
by a real particle resonantly interacting with two other particles.
This Letter shows how two impurities immersed in a Bose-Einstein condensate
can exhibit both phenomena. For a weak attraction with the condensate,
the two impurities form two polarons that interact through a weak
Yukawa attraction mediated by virtual excitations. For a resonant
attraction with the condensate, the exchanged excitation becomes a
real boson and the mediated interaction changes to a strong Efimov
attraction that can bind the two polarons. The resulting bipolarons
turn into in-medium Efimov trimers made of the two impurities and
one boson. Evidence of this physics could be seen in ultracold mixtures
of atoms.
\end{abstract}
\maketitle
A particle interacting with a surrounding medium can form a polaron,
i.e. it becomes dressed by a cloud of excitations of the medium that
alters its properties. This general concept, introduced by Landau
and Pekar~\cite{Landau1948} to describe electrons coupled to the
vibrations of a lattice in solids, has proved useful to understand
a variety of physical systems such as semi-conductors and superconductors~\cite{Devreese2009}.
In the last few years, polarons with arbitrarily strong interactions
with the medium could be investigated experimentally using ultra-cold
atoms~\cite{Schirotzek2009,Nascimbene2009,Kohstall2012,Koschorreck2012}.
These experiments have realised Fermi polarons (impurities embedded
in a Fermi sea) by mixing different kinds of fermionic atoms and tuning
their interaction by a Feshbach resonance. Recently, two experimental
works~\cite{Jorgensen2016,MingGuangHu2016} have reported the observation
of Bose polarons (impurities embedded in a Bose gas) by using bosonic
ultra-cold atoms. While the properties of a single Bose polaron are
interesting and theoretically challenging~\cite{Astrakharchik2004a,Cucchietti2006,Kalas2006,Bruderer2008,Tempere2009,Bei-Bing2009,Rath2013,Casteels2014,Shashi2014,Li2014,Volosniev2015,Grusdt2015,Vlietinck2015,Christensen2015,Levinsen2015,Ardila2015},
it is also of fundamental interest to understand the interaction between
Bose polarons induced by their medium~\cite{Ardila2015}. An exchange
of bosonic excitations is known to induce a Yukawa potential between
two polarons. This occurs in mixtures of bosonic and fermionic helium
liquids~\cite{Bardeen1967}. A similar phenomenon appears in high-energy
physics, where the nuclear force is mediated by mesons~\cite{Henley1962}.

On the other hand, at the few-body level, it is known that for sufficiently
strong interactions, an effective three-body force called the Efmov
attraction can bind three particles into one of infinitely many three-body
bound states, known as Efimov trimers~\cite{Efimov1970a,Braaten2006,Naidon2016}.
The Efimov attraction can be understood as an interaction between
two particles mediated by a third particle. It scales as the inverse
square of the distance between particles, conferring discrete scale
invariance to the system. Efimov trimers and their singular properties
have been observed in ultra-cold atom experiments in the last few
years~\cite{Braaten2007}, triggering the question of the influence
of the surrounding medium on these trimer states. In a condensate
of heavy bosons strongly interacting with light impurities, there
is a strong Efimov attraction that can form Efimov trimers of two
bosons and one impurity. Theoretical studies~\cite{Levinsen2015,Shchadilova2016}
have shown how a single polaron can turn into such an Efimov trimer,
and a similar effect was found for an impurity in a two-component
Fermi superfluid~\cite{Nishida2015,Yi2015}. Reference~\cite{Levinsen2015}
found that the in-medium Efimov trimer is stabilised by the surrounding
condensate. In the opposite limit of heavy impurities in a condensate
of light bosons, these trimers are very weak and the Efimov attraction
favours instead the formation of trimers of two impurities and one
boson. This indicates that in this case two polarons may turn into
one such Efimov trimer. A study~\cite{Zinner2013a} has suggested
that such an Efimov trimer would be weakened by the surrounding condensate.
 However, the theory could not completely describe the interaction
at large distance between the two impurities. The precise effect
of a surrounding Bose-Einstein condensate on Efimov trimers and the
mediated interaction thus remain to be clarified.

Motivated by these theoretical questions and the recent experiments
with ultra-cold atoms, this work presents a minimal description of
two impurities in a Bose-Einstein condensate that bridges the perturbative
regime of weakly attracted polarons and the non-perturbative regime
corresponding to a bound Efimov trimer immersed in a Bose-Einstein
condensate. In particular, the effective interaction between impurities
is shown to go from the Yukawa type, mediated by virtual bosonic excitations,
to the Efimov type, mediated by a real boson. This description is
based on the method of Refs.~\cite{Chevy2006,Levinsen2015} which
uses a variational wave function for the impurities and the excitations
of the medium. Here, the excitations are the Bogoliubov quasiparticles
of the condensate. In the following, at most one excitation will be
considered, which is the minimal requirement to reproduce the expected
Efimov three-body physics. First, the mediated interaction between
the two impurities will be derived, and then the energy spectrum of
the system will be presented and discussed. 

The Bose-Einstein condensate is assumed to be a homogeneous gas of
bosons of mass $m$ interacting via a weak pairwise interaction $U_{B}$,
whereas the interaction $U$ between an impurity and a boson may be
arbitrarily strong. No direct interaction between the impurities is
considered. The impurities are assumed to be identical bosons of mass
$M$. The Hamiltonian thus reads in second quantisation:
\begin{eqnarray}
\hat{H} & = & \sum_{\bm{k}}\epsilon_{k}b_{\bm{k}}^{\dagger}b_{\bm{k}}+\frac{1}{2V}\sum_{\bm{k},\bm{k}^{\prime},\bm{p}}U_{B}(\bm{p})b_{\bm{k^{\prime}-p}}^{\dagger}b_{\bm{k+p}}^{\dagger}b_{\bm{k}}b_{\bm{k}^{\prime}}\label{eq:Hamiltonian}\\
 &  & +\sum_{\bm{k}}\varepsilon_{k}c_{\bm{k}}^{\dagger}c_{\bm{k}}+\frac{1}{V}\sum_{\bm{k},\bm{k}^{\prime},\bm{p}}U(\bm{p})b_{\bm{k^{\prime}-p}}^{\dagger}c_{\bm{k}+\bm{p}}^{\dagger}c_{\bm{k}}b_{\bm{k}^{\prime}}\nonumber 
\end{eqnarray}
where $V$ is the system's volume, $\epsilon_{k}=\frac{\hbar^{2}k^{2}}{2m}$
and $b_{\bm{k}}$ are the kinetic energy and annihilation operator
for a boson with momentum $\bm{k}$, and $\varepsilon_{k}=\frac{\hbar^{2}k^{2}}{2M}$
and $c_{\bm{k}}$ are the kinetic energy and annihilation operator
for an impurity with momentum $\bm{k}$. Since the bosons are weakly
interacting, the first line of Eq.~(\ref{eq:Hamiltonian}) can be
approximately diagonalised as $\mathcal{E}_{0}+\sum_{\bm{k}}E_{k}\beta_{\bm{k}}^{\dagger}\beta_{\bm{k}}$,
by setting $b_{0}=\sqrt{N_{0}}$ and using for $\bm{k}\ne0$ the Bogoliubov
transformation
\begin{equation}
b_{\bm{k}}=u_{k}\beta_{\bm{k}}-v_{k}\beta_{-\bm{k}}^{\dagger},\label{eq:BogoliubovTransformation}
\end{equation}
where the operator $\beta_{\bm{k}}$ annihilates a quasi-particle
with momentum $\bm{k}$, $u_{k}^{2}=\frac{1}{2}\left(\frac{\epsilon_{k}+n_{0}U_{B}(0)}{E_{k}}+1\right)$,
$v_{k}^{2}=\frac{1}{2}\left(\frac{\epsilon_{k}+n_{0}U_{B}(0)}{E_{k}}-1\right)$,
and $E_{k}^{2}=\epsilon_{k}(\epsilon_{k}+2n_{0}U_{B}(0))$, where
$n_{0}=N_{0}/V$ is the condensate density. For convenience, the origin
of energy is set to the condensate ground-state energy $\mathcal{E}_{0}$.

The total wave function $\vert\Psi\rangle$ of the system can be expanded
exactly as a superposition of any number of excitations on top of
the Bose-Einstein consdensate $\vert\Phi\rangle$ and the two impurities.
Truncating this expansion to at most one excitation gives the following
ansatz, 
\begin{equation}
\vert\Psi\rangle=\left(\sum_{\bm{q}}\alpha_{\bm{q}}c_{\bm{q}}^{\dagger}c_{-\bm{q}}^{\dagger}+\sum_{\bm{q},\bm{q}^{\prime}}\alpha_{\bm{q},\bm{q}^{\prime}}c_{\bm{q}}^{\dagger}c_{\bm{q^{\prime}}}^{\dagger}\beta_{-\bm{q}-\bm{q^{\prime}}}^{\dagger}\right)\vert\Phi\rangle.\label{eq:Ansatz}
\end{equation}

Applying the variational principle $\langle\delta\Psi\vert H-E\vert\Psi\rangle=0$
to the Hamiltonian of Eq. (\ref{eq:Hamiltonian}) with the ansatz
of Eq.~(\ref{eq:Ansatz}), where $\alpha_{\bm{q}}$ and $\alpha_{\bm{q},\bm{q}^{\prime}}$
are varied independently, gives a set of two coupled equations:
\begin{eqnarray}
\left(2\varepsilon_{q}+2nU(0)-E\right)\alpha_{\bm{q}}\qquad\qquad\qquad\qquad\label{eq:Equation1}\\
+\frac{\sqrt{N_{0}}}{V}\sum_{\bm{k}}U(\bm{k})(u_{k}-v_{k})\left(\alpha_{\bm{q},\bm{k}-\bm{q}}+\alpha_{\bm{q}-\bm{k},-\bm{q}}\right) & = & 0,\nonumber 
\end{eqnarray}
\begin{multline}
\left(E_{k}+\varepsilon_{\vert\bm{k-q}\vert}+\varepsilon_{q}+2nU(0)-E\right)\alpha_{\bm{q},\bm{k-q}}\\
+\frac{1}{V}\sum_{\bm{p}}U(\bm{p}+\bm{k})\left(u_{k}u_{p}+v_{k}v_{p}\right)\left(\alpha_{\bm{q},-\bm{q}-\bm{p}}+\alpha_{\bm{q-p}-\bm{k},\bm{k}-\bm{q}}\right)\\
+\frac{\sqrt{N_{0}}}{V}U(\bm{k})(u_{k}-v_{k})\left(\alpha_{\bm{q}}+\alpha_{\bm{q}-\bm{k}}\right)=0,\label{eq:Equation2}
\end{multline}
where $n=n_{0}+\frac{1}{V}\sum_{\bm{k}}v_{k}^{2}\approx n_{0}(1+\frac{8}{3\sqrt{\pi}}\sqrt{n_{0}a_{B}^{3}})$
is the total density of bosons. Here, $a_{B}=\frac{m}{4\pi\hbar^{2}}U_{B}(0)$
is the boson scattering length in the Born approximation.

Let us first consider a weak interaction $U$, i.e. that can be treated
perturbatively. This imposes that the Born expansion of the scattering
length $a=a_{0}+a_{1}+\dots$ converges rapidly, and $a_{0}=\frac{2\mu}{4\pi\hbar^{2}}U(0)$
is much larger than $a_{1}=-\frac{2\mu}{4\pi\hbar^{2}}\frac{1}{V}\sum_{\bm{k}}\frac{U(k)^{2}}{\varepsilon_{k}+\epsilon_{k}}$,
where $\mu=(\frac{1}{M}+\frac{1}{m})^{-1}$ is the boson-impurity
reduced mass. In this case, one can neglect the sum in Eq.~(\ref{eq:Equation2}),
as it contributes to higher orders in $U$. Let us now consider the
limit $M\to\infty$ of heavy impurities separated by a vector $\bm{r}$,
and perform a Fourier transform with respect to $\bm{q}$, the conjugate
momentum of $\bm{r}$. Eliminating the second equation into the first,
one obtains
\begin{equation}
E^{\prime}=-2n_{0}\frac{1}{V}\sum_{\bm{k}}\frac{U(k)^{2}(u_{k}-v_{k})^{2}}{E_{k}-E^{\prime}}\left(1+e^{i\bm{k}\cdot\bm{r}}\right),\label{eq:PerturbativeBornOppenheimer}
\end{equation}
where $E^{\prime}=E-2nU(0)$. The solution $E(r)$ of this equation
as a function of $r$ gives the effective potential between the two
impurities in the Born-Oppenheimer approximation. Equation~(\ref{eq:PerturbativeBornOppenheimer})
shows that it decays as a Yukawa potential~(see Appendix A.1),
\begin{equation}
E(r)\stackrel[r\lesssim\xi]{\,}{=}E(\infty)-\frac{8\pi\hbar^{2}n_{0}}{2m}a_{0}^{2}\frac{\exp(-\sqrt{2}r/\xi)}{r},\label{eq:YukawaPotential}
\end{equation}
where $\xi=(8\pi n_{0}a_{B})^{-1/2}$ is the condensate coherence
length, and $E(\infty)=\frac{8\pi\hbar^{2}}{2\mu}\left(na_{0}+n_{0}\left(a_{1}+\sqrt{2}a_{0}^{2}/\xi\right)\right)$
is the asymptotic energy of the separated impurities, which is essentially
twice the mean-field energy $E_{MF}=\frac{4\pi\hbar^{2}}{2m}na$ of
a single impurity~\cite{Christensen2015}. This confirms well-known
results for small scattering lengths~\cite{Henley1962,Bardeen1967,Pethick2002}.
Note that, due its variational nature, the potential $E(r)$ is unreliable
for $r\gg\xi$, nevertheless it can be shown that the form of Eq.~(\ref{eq:YukawaPotential})
remains correct for $r\to\infty$~(see Appendix A.1 and A.2).

To investigate the non-perturbative regime, let us now consider the
limit of a contact interaction. It corresponds to a constant interaction
in momentum space, i.e. $U(\bm{k})=g<0$, up to some arbitrarily large
momentum cutoff $\Lambda$. The scattering length $a$ of this interaction
is given by the relation 
\begin{equation}
\frac{2\mu}{4\pi\hbar^{2}}\frac{1}{a}=\frac{1}{g}+\frac{1}{V}\sum_{\vert\bm{k}\vert<\Lambda}\frac{1}{\epsilon_{k}+\varepsilon_{k}},\label{eq:Renormalisation}
\end{equation}
which is used to renormalise all final results, i.e. express them
in terms of the scattering length $a$ instead of $g$. Using this
interaction in Eqs.~(\ref{eq:Equation1}-\ref{eq:Equation2}), one
encounters the terms $F_{\bm{q}}=g\frac{1}{V}\sum_{\bm{p}}u_{p}\alpha_{\bm{q},\bm{p}-\bm{q}}$
and $G_{\bm{q}}=g\frac{1}{V}\sum_{\bm{p}}v_{p}\alpha_{\bm{q},\bm{p}-\bm{q}}$.
Although $F_{\bm{q}}$ remains finite when $\Lambda\to\infty$, since
the sum in its expression diverges as $g^{-1}$ for a fixed value
of $a$, the term $G_{\bm{q}}$ vanishes, since the sum in its expression
does not diverge. In the end, one finds the following equation~(see
Appendix B.1),
\begin{equation}
\frac{F_{\bm{q}}}{T_{q}(E)}+\frac{1}{V}\sum_{\bm{k}}\frac{u_{k}^{2}\;F_{\bm{k}-\bm{q}}}{E_{k}+\varepsilon_{\vert\bm{k}-\bm{q}\vert}+\varepsilon_{q}-E}={\normalcolor \frac{2n_{0}}{E-2\varepsilon_{q}}}F_{\bm{q}},\label{eq:Contact3}
\end{equation}
where 
\begin{equation}
\frac{1}{T_{q}(E)}\!=\!\frac{2\mu}{4\pi\hbar^{2}}\frac{1}{a}+\frac{1}{V}\!\sum_{\bm{k}}\!\left(\!\frac{u_{k}^{2}}{E_{k}\!+\!\varepsilon_{\!\vert\bm{k}-\bm{q}\vert\!}\!+\!\varepsilon_{q}\!-\!E}-\frac{1}{\epsilon_{k}\!+\!\varepsilon_{k}}\!\right)\!\!.\label{eq:TwoBodyT}
\end{equation}
 As previously, one can find the mediated interaction in the Born-Oppenheimer
limit $M\to\infty$ by fixing the distance $r$ between the two impurities
and performing the Fourier transform of these equations. One obtains
\begin{equation}
\frac{2m}{4\pi\hbar^{2}}\frac{1}{a}+\frac{1}{V}\sum_{\bm{k}}\left(\frac{u_{k}^{2}}{E_{k}-E}(1+e^{-i\bm{k}\cdot\bm{r}})-\frac{1}{\epsilon_{k}}\right)=\frac{2n_{0}}{E}.\label{eq:ContactBornOppenheimer}
\end{equation}
This equation differs from Eq.~(4) of Ref.~\cite{Zinner2013a}
by its non-zero right-hand side and the coefficient $u_{k}^{2}\ne1$.
Let us consider its solution for weak $(1/a\to-\infty)$, unitary
$(1/a=0)$, and strong (1/$a\to+\infty$) boson-impurity interactions.

\begin{figure*}[t]
\begin{centering}
\includegraphics[bb=0bp 0bp 910bp 413bp,clip,scale=0.55]{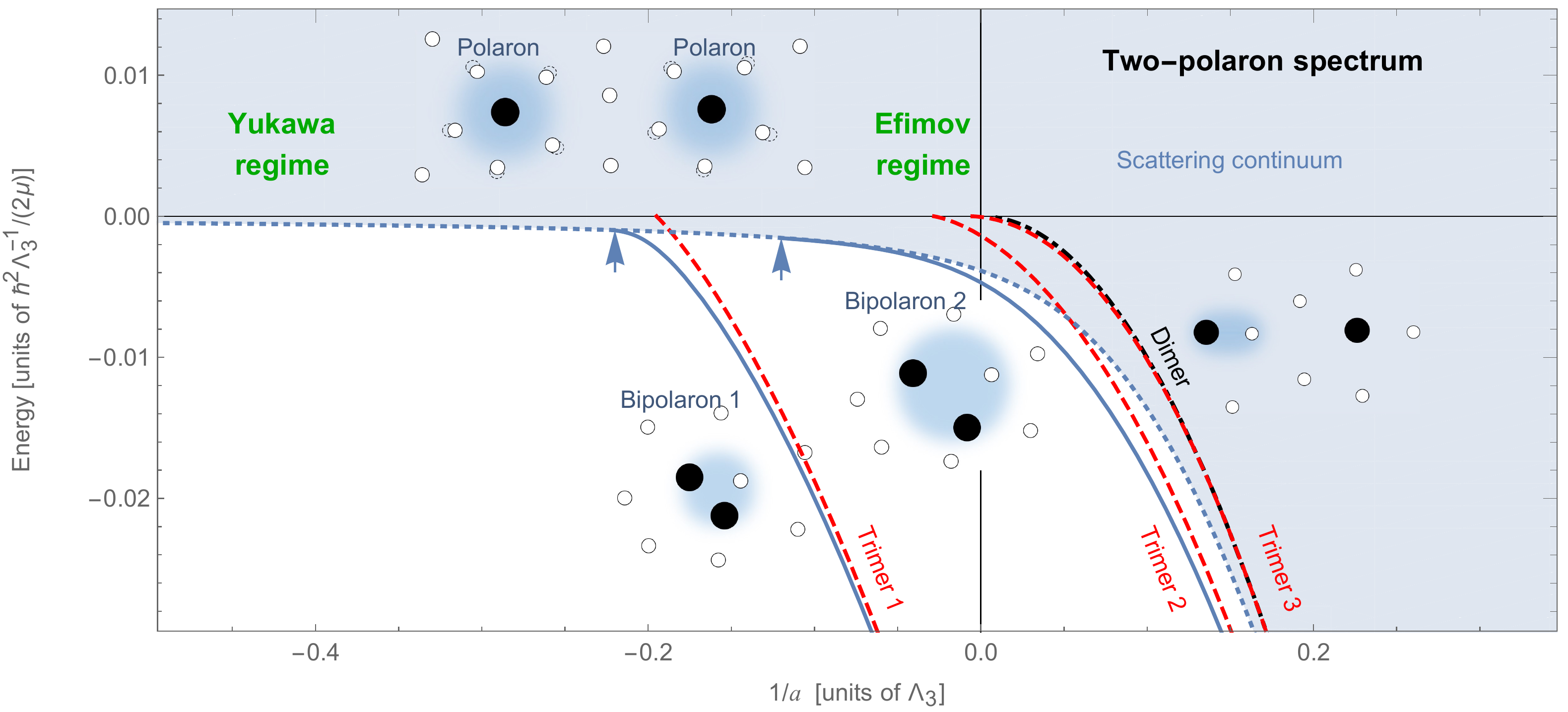}
\par\end{centering}
\caption{\label{fig:Energy-Spectrum}Energy spectrum of two polarons, i.e.
two impurities of mass $M$ in a condensate of bosons of mass $m=M/19$,
as a function of inverse scattering length between the impurities
and the bosons. The condensate density is $n_{0}=0.0005(\Lambda_{3})^{3}$,
and the boson scattering length is $a_{B}=1.5\Lambda_{3}^{-1}$, where
$\Lambda_{3}$ is a three-body cutoff inversely proportional to the
range of the boson-impurity interaction. The spectrum is obtained
from Eq.~(\ref{eq:Contact3}), which is expected to be quantitative
for $1/a\lesssim0$. The shaded area represents the scattering continuum
of the two attractive polarons and its threshold is shown as a dotted
curve. The solid curves correspond to the bound states (bipolarons).
The points where they appear from the polaron scattering threshold
are indicated by the vertical arrows. For reference, the black dot-dashed
curve shows the boson-impurity dimer energy in vacuum, and the red
dashed curves correspond to the boson-impurity-impurity trimer energies
in vacuum. }
\end{figure*}

For weak boson-impurity interactions ($a<0$ and $\vert a\vert\ll a_{B}$),
the solution $E(r)$ of Eq.~(\ref{eq:ContactBornOppenheimer}) is
of the Yukawa type~(see Appendix B.2),
\begin{equation}
E(r)\stackrel[r\lesssim\xi]{\,}{=}E(\infty)-\frac{8\pi\hbar^{2}n_{0}}{2m}a^{2}\frac{\exp(-\frac{4+\pi}{2\sqrt{2\pi}}r/\xi)}{r},\label{eq:ContactSmallEAsymptote}
\end{equation}
where $E(\infty)=\frac{8\pi\hbar^{2}}{2m}n_{0}a\left(1+\frac{4+\pi}{2\sqrt{2\pi}}a/\xi\right)\approx2E_{\text{MF}}$.
It can be seen that this potential is slightly different from the
perturbative result of Eq.~(\ref{eq:YukawaPotential}). This is attributed
to the limitation of the variational ansatz, which only provides an
upper bound of the exact potential~(see Appendix B.2). 

Let us now consider the cases $1/a\to0$ and $1/a\to+\infty$. For
sufficiently large $\vert E\vert=\frac{\hbar^{2}}{2m}\kappa^{2}\gg\frac{\hbar^{2}}{2m}\xi^{-2}$,
Eq.~(\ref{eq:ContactBornOppenheimer}) can be approximated to first
order in $a_{B}$ as
\begin{equation}
\frac{1}{a}-\kappa-\frac{4\pi}{\kappa}n_{0}a_{B}+\left(\frac{1}{r}-\frac{4\pi n_{0}a_{B}}{\kappa}\right)e^{-\kappa r}=-\frac{8\pi n_{0}}{\kappa^{2}}.\label{eq:ContactLargeE}
\end{equation}
It follows that in the large scattering length limit $1/a\to0$, the
mediated interaction $E(r)$ has the form~(see Appendix B.2)
\begin{equation}
\begin{cases}
-\frac{\hbar^{2}}{2m}\frac{W(1)^{2}}{r^{2}} & \mbox{for }r\ll L\quad\text{\quad\quad(a)}\\
\qquad & \qquad\\
-\frac{\hbar^{2}\kappa_{\infty}^{2}}{2m}\left(1+\frac{2}{3}\left(\frac{L}{r}-\frac{L^{2}}{2\xi^{2}}\right)e^{-\kappa_{\infty}r}\right) & \mbox{for }L\lesssim r\ll\xi\quad\text{(b)}
\end{cases}\label{eq:ContactLargeEAsymptote}
\end{equation}
where $W(1)\approx0.567$, $\kappa_{\infty}=L^{-1}-L/(6\xi^{2})\sim L^{-1}$,
and $L=(8\pi n_{0})^{-1/3}$. One recognises at short distances the
$1/r^{2}$ Efimov attraction (in the Born-Oppenheimer limit~\cite{Naidon2016})
between two impurities mediated by a boson. The Efimov attraction
can support an infinite number of bound states. However, here it is
truncated at distances on the order of the mean boson spacing $L=(8\pi n_{0})^{-1/3}$,
and asymptotes to the energy $E(\infty)\approx-\frac{\hbar^{2}}{2mL^{2}}$.
As a result, the infinite number of possible trimer states in vacuum
is reduced to a finite number, such that only those trimers whose
energy is lower or comparable to $E(\infty)$ survive in the presence
of the condensate.

Finally, as the boson-impurity interaction is strengthened towards
small positive scattering length $a$, each polaron is expected to
turn into a dimer of energy $E_{\text{d}}=-\frac{\hbar^{2}}{2ma^{2}}$,
as each impurity should strongly bind with a nearby boson. However,
in the present theory, the asymptotic energy $E(\infty)$ of the two
separated impurities goes to $E_{\text{d}}$ instead of $2E_{\text{d}}$,
as can be seen from Eq.~(\ref{eq:ContactLargeE}) for $r\to\infty$
and $1/a\to+\infty$. The reason is that the ansatz of Eq.~(\ref{eq:Ansatz})
includes only one bosonic excitation, and as a result only one impurity
can bind with that excitation. A more quantitative treatment of the
scattering threshold of the impurities in this regime would thus require
at least two bosonic excitations. 

Let us now turn to the energy spectrum of the system. In the contact
model, the Efimov attraction exists at infinitely small distances,
as seen in Eq.~(\ref{eq:ContactLargeEAsymptote}a), leading to the
so-called Thomas collapse~\cite{Thomas1935,Braaten2006}. Some additional
short-range scale is necessary to cure this problem and set the three-body
observables~\cite{Naidon2016}. This can be done at the two-body
level by keeping a finite momentum cutoff $\Lambda$ for the sum in
Eq.~(\ref{eq:Contact3}). Alternatively, one may introduce a three-body
force. The simplest way to introduce such a force is to set a momentum
cutoff $\Lambda_{3}$ on the second Jacobi momentum, i.e. the argument
of $F$ in Eq.~(\ref{eq:Contact3}). In atomic gases, this three-body
parameter is related to the van der Waals length of the atoms~\cite{Wang2012,Naidon2014a,Wang2014,Naidon2016}.
Figure~\ref{fig:Energy-Spectrum} represents the exact energy spectrum
of the system for a mass ratio $M/m=19$ (such as caesium-133 atoms
in a lithium-7 condensate), as a function of $1/a$, and calculated
numerically from Eq.~(\ref{eq:Contact3}) with a three-body cutoff
$\Lambda_{3}$.

For any boson-impurity interaction, the spectrum shows a continuum
corresponding to scattering states of two attractive polarons. Its
threshold, shown by the dotted curve in Fig.~\ref{fig:Energy-Spectrum},
corresponds to the asymptotic limit of the mediated interaction, which
for the large mass ratio used here is well approximated by the Born-Oppenheimer
threshold given by the solution of Eq.~(\ref{eq:ContactLargeE})
for $r\to\infty$. As noted before, the threshold corresponds to the
mean-field energy $2E_{\text{MF}}$ of two polarons for small $a<0$,
and (unphysically) asymptotes to the energy $E_{\text{d}}$ of a single
dimer for small $a>0$. The spectrum also features discrete bound
states for sufficiently strong boson-impurity interaction. This is
expected since the mediated interaction becomes strong enough to bind
the two polarons into bipolarons as it gradually turns from a weak
Yukawa potential into a strong Efimov attraction. As the interaction
further increases, the bipolarons (shown as solid curves in Fig.~\ref{fig:Energy-Spectrum})
turn into Efimov trimers made of two impurities and one boson. As
anticipated from the Born-Oppenheimer potential between the two impurities
- see Eq.~(\ref{eq:ContactLargeEAsymptote}) - only the trimers whose
energy is lower than the polaron scattering threshold survive in the
presence of the condensate. Near unitarity ($1/a=0$), the bipolaron
energies are pushed down from the trimer energies in vacuum due the
attractive effect of the surrounding bosons, but the binding energies
relative to the polaron scattering threshold are smaller than in vacuum.
Near unitarity, the trimers are therefore \emph{weakened by the condensate}.
However, interestingly, the bipolarons exist for weaker boson-impurity
interactions than the vacuum trimers. The scattering lengths at which
the bipolarons appear (indicated by arrows in Fig.~\ref{fig:Energy-Spectrum})
are indeed reduced in magnitude with respect to vacuum. In this sense,
the condensate \emph{favours the appearance of the trimers}. This
is especially true when the polaron scattering threshold at unitarity
($\sim-\hbar^{2}/(2\mu L^{2})$) is comparable to the energy of an
Efimov trimer in vacuum, as shown by the second bipolaron in Fig.~\ref{fig:Energy-Spectrum}.
It should be mentioned that the boson-boson interaction has the opposite
effect of weakening the bipolarons, but this effect remains small
in the assumed dilute regime $na_{B}^{3}\ll1$.

Finally, it is important to specify the range of validity of the present
treatement. As noted above, for sufficiently large attraction between
the impurities and the bosons, more than one bosonic excitation are
needed. Since the bipolarons are seen in Fig.~\ref{fig:Energy-Spectrum}
to correlate to the vacuum trimer states of two impurities and one
boson, it is natural to expect that an additional bosonic excitation
would correlate them to tetramer states of two impurities and two
bosons. Such tetramers do exist for weak interactions between the
bosons and would significantly affect the two-polaron spectrum beyond
the results presented here. However, for a moderate interaction between
the bosons $a_{B}\gtrsim\Lambda_{3}^{-1}$ that is typical for ultracold
atoms, the tetramers are found to be suppressed for $1/a\lesssim0$~\cite{Naidon2018}.
It is therefore expected that the present theory is quantitative in
this regime.

In summary, a simple variational ansatz has been used to investigate
the problem of two impurities in a Bose-Einstein condensate. The ansatz
bridges the well-known perturbative regime to the non-perturbative
regime, where the Bose-mediated interaction takes the form of the
Efimov attraction. It shows that the two polarons formed by the two
impurities merge into one or several Efimov trimers for sufficiently
strong interaction. The stability of these bipolarons under the influence
of the condensate has also been revealed. Although their binding energy
is reduced near unitarity with respect to that of trimers in vacuum,
they exist for smaller interaction as the density of the condensate
is increased. In a mixture of resonantly interacting ultra-cold atoms,
this would appear as a boson-density-dependent shift of the three-body
loss peaks associated with the appearance of Efimov trimers. The direct
effect of the mediated interaction between impurities could be observed
as an impurity-density-dependent mean-field shift (estimated on the
order of a few percents) in the single-polaron energy spectrum.\\

The author thanks Nguyen Thanh Phuc, Takumi Doi, and Tetsuo Hatsuda
for helpful discussions. This work was partially supported by the
RIKEN Incentive Research, iTHES, and iTHEMS projects..

\bibliographystyle{apsrev4-1}

\begin{widetext}

\section*{Appendix}

\subsection{Derivation of the Yukawa potential}

\subsubsection{Derivation within the variational ansatz}

The Yukawa potential of Eq.~(\ref{eq:YukawaPotential}) is obtained
as follows. For sufficiently small scattering length, the term $E^{\prime}$
in the denominator of Eq.~(\ref{eq:PerturbativeBornOppenheimer})
may be neglected as it contributes to higher orders, and the term
$(u_{k}-v_{k})^{2}=\epsilon_{k}/E_{k}$. It follows that: 
\begin{align}
E^{\prime} & =-2n_{0}\frac{1}{V}\sum_{\bm{k}}\frac{U(k)^{2}\epsilon_{k}}{E_{k}^{2}}\left(1+e^{i\bm{k}\cdot\bm{r}}\right)\label{eq:A1}\\
 & =-2n_{0}\frac{1}{V}\sum_{\bm{k}}\frac{U(k)^{2}}{\epsilon_{k}+2n_{0}U_{B}(0)}\left(1+e^{i\bm{k}\cdot\bm{r}}\right)\label{eq:A2}
\end{align}

Using $\epsilon_{k}=\hbar^{2}k^{2}/(2m)$ and $U_{B}(0)=4\pi\hbar^{2}a_{B}/m$,
one gets

\begin{equation}
E^{\prime}=\underbrace{-2n_{0}\frac{2m}{\hbar^{2}}\frac{1}{V}\sum_{\bm{k}}\frac{U(k)^{2}}{k^{2}+2\xi^{-2}}}_{E^{\prime}(\infty)}-\underbrace{2n_{0}\frac{2m}{\hbar^{2}}\frac{1}{V}\sum_{\bm{k}}\frac{U(k)^{2}}{k^{2}+2\xi^{-2}}e^{i\bm{k}\cdot\bm{r}}}_{E^{\prime}(r)-E^{\prime}(\infty)}\label{eq:A3}
\end{equation}
with $\xi=1/\sqrt{8\pi n_{0}a_{B}}$. The term $E^{\prime}(\infty)$
does not depend on $r$ and can be written as a follows:

\begin{align*}
E^{\prime}(\infty) & =-2n_{0}\frac{2m}{\hbar^{2}}\left(\frac{1}{V}\sum_{\bm{k}}\frac{U(k)^{2}}{k^{2}}+\frac{1}{V}\sum_{\bm{k}}U(k)^{2}\left(\frac{1}{k^{2}+2\xi^{-2}}-\frac{1}{k^{2}}\right)\right)
\end{align*}

The first sum in the above expression converges due to the decay of
$U(k)$ at large $k$, and is related to the second term $a_{1}=-\frac{2m}{4\pi\hbar^{2}}\frac{1}{V}\sum_{\bm{k}}\frac{U(k)^{2}}{\epsilon_{k}}$
in the Born expansion of the scattering length $a$. In the second
sum, one may take $U(k)\approx U(0)=\frac{4\pi\hbar^{2}}{2m}a_{0}$
since the momentum range of $U$ is typically much larger than $\xi^{-1}$.
One then obtains

\begin{align}
E^{\prime}(\infty) & =-2n_{0}\frac{2m}{\hbar^{2}}\left(-4\pi\left(\frac{\hbar^{2}}{2m}\right)^{2}a_{1}+U(0)^{2}\frac{1}{V}\sum_{\bm{k}}\left(\frac{1}{k^{2}+2\xi^{-2}}-\frac{1}{k^{2}}\right)\right)\nonumber \\
 & =-2n_{0}\frac{2m}{\hbar^{2}}\left(-4\pi\left(\frac{\hbar^{2}}{2m}\right)^{2}a_{1}+4\pi\left(\frac{\hbar^{2}}{2m}\right)^{2}a_{0}^{2}\underbrace{\frac{2}{\pi}\int_{0}^{\infty}\left(\frac{k^{2}}{k^{2}+2\xi^{-2}}-1\right)dk}_{-\sqrt{2}/\xi}\right)\nonumber \\
 & =\frac{8\pi n_{0}\hbar^{2}}{2m}\left(a_{1}+\sqrt{2}a_{0}^{2}/\xi\right)\label{eq:A4}
\end{align}

The last sum in Eq.~(\ref{eq:A3}) goes to zero as $r\to\infty$.
Its asymptotic behaviour at large $r$ may be obtained from the low-momentum
contribution in the sum. In this limit, $U(k)$ may be approximated
by $U(0)$, i.e. 
\[
E^{\prime}(r)-E^{\prime}(\infty)\xrightarrow[r\to\infty]{}-2n_{0}\frac{2m}{\hbar^{2}}U(0)^{2}\times\frac{1}{V}\sum_{\bm{k}}\frac{1}{k^{2}+2\xi^{-2}}e^{i\bm{k}\cdot\bm{r}}
\]

One recognises in the sum the Fourier transform of $\exp(-\sqrt{2}r/\xi)/(4\pi r)$,
and using the relation $U(0)=\frac{4\pi\hbar^{2}}{2\mu}a_{0}$, one
finally gets
\begin{equation}
E^{\prime}(r)-E^{\prime}(\infty)\xrightarrow[r\to\infty]{}-\frac{4\pi\hbar^{2}n_{0}}{m}a_{0}^{2}\frac{\exp(-\sqrt{2}r/\xi)}{r},\label{eq:A5}
\end{equation}
which establishes Eq.~(\ref{eq:YukawaPotential}).

\subsubsection{Shortcoming of the variational ansatz}

We should note that while Eq.~(\ref{eq:A5}) gives the form of $E^{\prime}(r)$
in the limit of small interaction, the convergence to this potential
is not uniform. For small but finite attraction $U$, there is indeed
a distance beyond which the term $E^{\prime}$ in the denominator
of Eq.~(\ref{eq:PerturbativeBornOppenheimer}) may not be neglected.
Taking into account this term, one can find the true asymptotic form
to be
\[
E^{\prime}(r)-E^{\prime}(\infty)\xrightarrow[r\to\infty]{}-2n_{0}\frac{1}{V}\sum_{\bm{k}}\frac{U(k)^{2}\epsilon_{k}}{E_{k}(E_{k}-E^{\prime}(\infty))}e^{i\bm{k}\cdot\bm{r}}\approx-2n_{0}U(0)^{2}\frac{2m}{4\pi\hbar^{2}}\frac{2}{\pi}\frac{1}{r}\int_{0}^{\infty}F(k)\sin krdk
\]
where $F(k)=k^{2}/\left(\sqrt{k^{2}+2\xi^{-2}}\left(k\sqrt{k^{2}+2\xi^{-2}}-\frac{2m}{\hbar^{2}}E^{\prime}(\infty)\right)\right)$.
Using the property $\int_{0}^{\infty}F(k)\sin krdk\xrightarrow[r\to\infty]{}F(0)/r-F^{(2)}(0)/r^{3}+O(1/r^{5})$,
and the facts that $F(0)=0$ and $F^{(2)}(0)=-\sqrt{2}\xi/\left(\frac{2m}{\hbar^{2}}E^{\prime}(\infty)\right)$,
one obtains
\[
E^{\prime}(r)-E^{\prime}(\infty)\xrightarrow[r\to\infty]{}2n_{0}\frac{U(0)^{2}}{2\pi^{2}}\frac{\sqrt{2}\xi}{\vert E^{\prime}(\infty)\vert}\frac{1}{r^{4}}.
\]
One can show that this $1/r^{4}$ repulsion occurs for $r\gg\left(\xi\frac{2m}{\hbar^{2}}E^{\prime}(\infty)\right)^{-1}$.
However, this asymptotic behaviour has no physical reality as it is
an artifact of the variational ansatz of Eq.~(\ref{eq:Ansatz}) for
large distances. Indeed, the variational ansatz only gives an upper
bound of the exact potential and its scattering threshold. Even if
the variational threshold is only slightly above the exact one, the
way the variational potential asymptotes to this threshold may be
completely different from the way the exact potential asymptotes to
the exact threshold. As a result, the analytic form of the asymptote
may be wrong, as seen here.

\subsubsection{Exact derivation}

It turns out that the perturbative result Eq.~(\ref{eq:A5}) holds
exactly beyond the variational ansatz of Eq.~(\ref{eq:Ansatz}) and
to infinite distances. To show this, let us first treat the boson-impurity
as a perturbation to first order in the Hamiltonian. We obtain the
following Frölich-like Hamiltonian:
\[
H=\sum_{\bm{k}}E_{k}\beta_{\bm{k}}^{\dagger}\beta_{\bm{k}}+\sum_{\bm{k}}\left(\varepsilon_{\bm{k}}+nU(0)\right)c_{\bm{k}}^{\dagger}c_{\bm{k}}+\sum_{\bm{k},\bm{p}}g_{\bm{k}}(\beta_{-\bm{k}}^{\dagger}+\beta_{\bm{k}})c_{\bm{p}+\bm{k}}^{\dagger}c_{\bm{p}},
\]
where $g_{k}=U(k)\sqrt{n_{0}}(u_{k}-v_{k})$. Taking the limit of
static impurities ($M\to\infty$), we get $\varepsilon_{k}\to0$ and
write $\sum_{\bm{p}}c_{\bm{p}+\bm{k}}^{\dagger}c_{\bm{p}}=\int d^{3}\bm{R}n_{c}(\bm{R})e^{i\bm{k}\cdot\bm{R}}$,
where $n_{c}(\bm{R})$ is the density of the impurities. For two impurities
separated by $r$, we have $n_{c}(\bm{R})=\delta^{3}(\bm{R})+\delta^{3}(\bm{R}+\bm{r})$.
This gives the $\bm{r}$-dependent Hamiltonian,
\[
H^{\prime}(\bm{r})=H(\bm{r})-2nU(0)=\sum_{\bm{k}}\left[E_{k}\beta_{\bm{k}}^{\dagger}\beta_{\bm{k}}+g_{\bm{k}}\left(\beta_{\bm{k}}^{\dagger}\left(1+e^{i\bm{k}\cdot\bm{r}}\right)+\beta_{\bm{k}}\left(1+e^{-i\bm{k}\cdot\bm{r}}\right)\right)\right].
\]
We now note that the Hamiltonian can be diagonalised exactly by introducing
the operator $\tilde{\beta}_{\bm{k}}$ such that
\[
\beta_{\bm{k}}=\tilde{\beta}_{\bm{k}}-\frac{g_{k}}{E_{k}}\left(1+e^{i\bm{k}\cdot\bm{R}}\right).
\]

One can check that this operator satisfies the bosonic commutation
relations $[\tilde{\beta}_{\bm{k}},\tilde{\beta}_{\bm{q}}]=0$ and
$[\tilde{\beta}_{\bm{k}},\tilde{\beta}_{\bm{q}}^{\dagger}]=\delta_{\bm{k},\bm{q}}$,
and the Hamiltonian expressed in terms of this operator reads
\[
H^{\prime}(\bm{R})=-2\sum_{\bm{k}}\frac{g_{k}^{2}}{E_{k}}\left(1+e^{i\bm{k}\cdot\bm{R}}\right)+\sum_{\bm{k}}E_{k}\tilde{\beta}_{\bm{k}}^{\dagger}\tilde{\beta}_{\bm{k}}
\]

The ground state is therefore given by 
\[
E^{\prime}(R)=-2n_{0}\sum_{\bm{k}}\frac{(u_{k}-v_{k})^{2}U(k)^{2}}{E_{k}}\left(1+e^{i\bm{k}\cdot\bm{R}}\right)
\]
which is exactly the same as Eq.~(\ref{eq:A1}). It follows that
the results Eqs.~(\ref{eq:A4}) and (\ref{eq:A5}) are exact in the
perturbative limit, beyond the variational ansatz Eq.~(\ref{eq:Ansatz}).

\subsection{Equation and solution for the contact model}

\subsubsection{Derivation of the equation}

The following provides the derivation of Eq.~(\ref{eq:Contact3}).

Starting from the general equations Eqs.~(\ref{eq:Equation1}-\ref{eq:Equation2}),
one performs the changes $\alpha_{\bm{q}-\bm{k},-\bm{q}}=\alpha_{-\bm{q},\bm{q}-\bm{k}}$
in Eq.~(\ref{eq:Equation1}) and $\alpha_{\bm{q-p}-\bm{k},\bm{k}-\bm{q}}=\alpha_{\bm{k}-\bm{q},\bm{q-p}-\bm{k}}$
in Eq.~(\ref{eq:Equation2}), using the bosonic exchange symmetry
$\alpha_{\bm{q},\bm{q}^{\prime}}=\alpha_{\bm{q}^{\prime},\bm{q}}$.
Then, one sets the potential $U(\bm{k})=g$ for $k<\Lambda$, $U(\bm{k})=0$
for $k\ge\Lambda$. This yields
\begin{equation}
\left(2\varepsilon_{q}-E^{\prime}\right)\alpha_{\bm{q}}+g\frac{\sqrt{N_{0}}}{V}\sum_{\bm{k}}^{k<\Lambda}(u_{k}-v_{k})\left(\alpha_{\bm{q},\bm{k}-\bm{q}}+\alpha_{-\bm{q},\bm{q}-\bm{k}}\right)=0\label{eq:B1}
\end{equation}
\begin{equation}
\left(E_{k}+\varepsilon_{\vert\bm{k-q}\vert}+\varepsilon_{q}-E^{\prime}\right)\alpha_{\bm{q},\bm{k-q}}+g\frac{1}{V}\sum_{\bm{p}}^{\vert\bm{p}+\bm{k}\vert<\Lambda}\left(u_{k}u_{p}+v_{k}v_{p}\right)\left(\alpha_{\bm{q},-\bm{q}-\bm{p}}+\alpha_{\bm{k}-\bm{q},\bm{q-p}-\bm{k}}\right)+g\frac{\sqrt{N_{0}}}{V}(u_{k}-v_{k})\left(\alpha_{\bm{q}}+\alpha_{\bm{q}-\bm{k}}\right)=0,\label{eq:B2}
\end{equation}

In the first equation, one can change the term $\alpha_{-\bm{q},\bm{q}-\bm{k}}$
into $\alpha_{-\bm{q},\bm{k}+\bm{q}}$ by performing the change of
variable $\bm{k}\to-\bm{k}$. The equation then reads,
\begin{equation}
\left(2\varepsilon_{q}-E^{\prime}\right)\alpha_{\bm{q}}+\sqrt{N_{0}}\left(F_{\bm{q}}+F_{-\bm{q}}-G_{\bm{q}}-G_{-\bm{q}}\right)=0\label{eq:B3}
\end{equation}
where the terms $F_{\bm{q}}$ and $G_{\bm{q}}$ are defined by
\begin{equation}
F_{\bm{q}}=g\frac{1}{V}\sum_{\bm{k}}^{k<\Lambda}u_{k}\alpha_{\bm{q},\bm{k}-\bm{q}}\label{eq:B4}
\end{equation}
\begin{equation}
G_{\bm{q}}=g\frac{1}{V}\sum_{\bm{k}}^{k<\Lambda}v_{k}\alpha_{\bm{q},\bm{k}-\bm{q}}\label{eq:B5}
\end{equation}

Next, the change of variable $\bm{p}\to-\bm{p}$ is performed in Eq.~(\ref{eq:B2}).
For sufficiently large $\Lambda$, the sum $\sum_{\bm{p}}^{\vert\bm{p}-\bm{k}\vert<\Lambda}$
can be approximated by $\sum_{\bm{p}}^{p<\Lambda}$, so that Eq.~(\ref{eq:B2})
can be expressed in terms of $F$ and $G$:
\begin{equation}
\left(E_{k}+\varepsilon_{\vert\bm{k-q}\vert}+\varepsilon_{q}-E^{\prime}\right)\alpha_{\bm{q},\bm{k-q}}+u_{k}(F_{\bm{q}}+F_{\bm{k}-\bm{q}})+v_{k}(G_{\bm{q}}+G_{\bm{k}-\bm{q}})+g\frac{\sqrt{N_{0}}}{V}(u_{k}-v_{k})\left(\alpha_{\bm{q}}+\alpha_{\bm{q}-\bm{k}}\right)=0\label{eq:B6}
\end{equation}

Using this equation to express $\alpha_{\bm{q},\bm{k}-\bm{q}}$ in
Eqs.~(\ref{eq:B4}) and (\ref{eq:B5}), one finds
\begin{equation}
\frac{1}{g}F_{\bm{q}}=-\frac{1}{V}\sum_{\bm{k}}^{k<\Lambda}u_{k}\frac{u_{k}(F_{\bm{q}}+F_{\bm{k}-\bm{q}})+v_{k}(G_{\bm{q}}+G_{\bm{k}-\bm{q}})+g\frac{\sqrt{N_{0}}}{V}(u_{k}-v_{k})\left(\alpha_{\bm{q}}+\alpha_{\bm{q}-\bm{k}}\right)}{E_{k}+\varepsilon_{\vert\bm{k-q}\vert}+\varepsilon_{q}-E^{\prime}}\label{eq:B7}
\end{equation}
\begin{equation}
\frac{1}{g}G_{\bm{q}}=-\frac{1}{V}\sum_{\bm{k}}^{k<\Lambda}v_{k}\frac{u_{k}(F_{\bm{q}}+F_{\bm{k}-\bm{q}})+v_{k}(G_{\bm{q}}+G_{\bm{k}-\bm{q}})+g\frac{\sqrt{N_{0}}}{V}(u_{k}-v_{k})\left(\alpha_{\bm{q}}+\alpha_{\bm{q}-\bm{k}}\right)}{E_{k}+\varepsilon_{\vert\bm{k-q}\vert}+\varepsilon_{q}-E^{\prime}}\label{eq:B8}
\end{equation}

Owing to the renormalisation relation Eq.~(\ref{eq:Renormalisation}),
for a fixed scattering length $a$, the term $1/g$ in the left-hand
side of Eqs. (\ref{eq:B7}) and (\ref{eq:B8}) diverges as $\Lambda$
for very large $\Lambda$. In Eq.~(\ref{eq:B7}), this divergence
in the left-hand side is cancelled by another divergent term in the
right-hand side. In contrast, in Eq.~(\ref{eq:B8}), the right-hand
side does not diverge for large $\Lambda$. Indeed, for large $\bm{k}$,
the denominator in Eq.~(\ref{eq:B8}) is $\sim k^{2}$, and the numerator
involves the terms $v_{k}u_{k}\sim k^{-2}$ and $v_{k}^{2}\sim k^{-4}$.
The term in the sum of Eq.~(\ref{eq:B8}) thus decay as $k^{-4}$
or faster, and the sum is therefore convergent. One concludes that
$G$ may be neglected for sufficiently large $\Lambda$.

There only remain two equations, Eq.~(\ref{eq:B3}) and (\ref{eq:B7}),
which for $G=0$ read 
\begin{equation}
\left(2\varepsilon_{q}-E^{\prime}\right)\alpha_{\bm{q}}+\sqrt{N_{0}}\left(F_{\bm{q}}+F_{-\bm{q}}\right)=0\label{eq:B9}
\end{equation}
\begin{equation}
\frac{1}{g}F_{\bm{q}}=-\frac{1}{V}\sum_{\bm{k}}^{k<\Lambda}\frac{u_{k}^{2}(F_{\bm{q}}+F_{\bm{k}-\bm{q}})+g\frac{\sqrt{N_{0}}}{V}u_{k}(u_{k}-v_{k})\left(\alpha_{\bm{q}}+\alpha_{\bm{q}-\bm{k}}\right)}{E_{k}+\varepsilon_{\vert\bm{k-q}\vert}+\varepsilon_{q}-E^{\prime}}\label{eq:B10}
\end{equation}

One can further rewrite Eq.~(\ref{eq:B10}) as
\begin{align}
\left(\frac{1}{g}+\frac{1}{V}\sum_{\bm{k}}^{k<\Lambda}\frac{u_{k}^{2}}{E_{k}+\varepsilon_{\vert\bm{k-q}\vert}+\varepsilon_{q}-E^{\prime}}\right)F_{\bm{q}}= & -\frac{1}{V}\sum_{\bm{k}}^{k<\Lambda}\frac{u_{k}^{2}}{E_{k}+\varepsilon_{\vert\bm{k-q}\vert}+\varepsilon_{q}-E^{\prime}}F_{\bm{k}-\bm{q}}\label{eq:B11}\\
 & -g\frac{\sqrt{N_{0}}}{V}\left[\frac{1}{V}\sum_{\bm{k}}^{k<\Lambda}\frac{u_{k}(u_{k}-v_{k})}{E_{k}+\varepsilon_{\vert\bm{k-q}\vert}+\varepsilon_{q}-E^{\prime}}\right]\alpha_{\bm{q}}\nonumber \\
 & -g\frac{\sqrt{N_{0}}}{V}\left[\frac{1}{V}\sum_{\bm{k}}^{k<\Lambda}\frac{u_{k}(u_{k}-v_{k})\alpha_{\bm{q}-\bm{k}}}{E_{k}+\varepsilon_{\vert\bm{k-q}\vert}+\varepsilon_{q}-E^{\prime}}\right]\nonumber 
\end{align}

The sum in the last term is convergent, since for large $\bm{k}$,
$u_{k}(u_{k}-v_{k})\sim1$ and $\alpha_{\bm{q}-\bm{k}}\lesssim k^{-2}$
according to Eq.~(\ref{eq:B9}). Since it is multiplied by the vanishing
factor $g$, it can therefore be neglected. On the other hand, the
sum in the second term of Eq.~(\ref{eq:B11}) diverges as $-1/g$
- as seen from Eq.~(\ref{eq:Renormalisation}). It therefore cancels
the factor $g$ for large enough $\Lambda$. Finally, the divergence
of the sum in the left-hand side of Eq.~(\ref{eq:B11}) is cancelled
by the term $1/g$, which can be done explicitly by using the renormalisation
relation Eq.~(\ref{eq:Renormalisation}). Finally, Eq.~(\ref{eq:B11})
simplifies to
\begin{equation}
\frac{1}{T_{\bm{q}}(E^{\prime})}F_{\bm{q}}+\frac{1}{V}\sum_{\bm{k}}^{k<\Lambda}\frac{u_{k}^{2}}{E_{k}+\varepsilon_{\vert\bm{k-q}\vert}+\varepsilon_{q}-E^{\prime}}F_{\bm{k}-\bm{q}}=\frac{\sqrt{N_{0}}}{V}\alpha_{\bm{q}}\label{eq:B12}
\end{equation}
with
\begin{equation}
\frac{1}{T_{q}(E^{\prime})}=\frac{2\mu}{4\pi\hbar^{2}}\frac{1}{a}+\frac{1}{V}\sum_{\bm{k}}^{k<\Lambda}\left(\frac{u_{k}^{2}}{E_{k}+\varepsilon_{\vert\bm{k}-\bm{q}\vert}+\varepsilon_{q}-E^{\prime}}-\frac{1}{\epsilon_{k}+\varepsilon_{k}}\right).\label{eq:B13}
\end{equation}
The energy $E^{\prime}=E-2ng$ may be replaced by $E$ since $g$
vanishes for large $\Lambda$. As a last step, one can perform the
changes $\bm{q}\to-\bm{q}$ and $\bm{k}\to-\bm{k}$ in Eq.~(\ref{eq:B12}),
and observe that $F_{-\bm{q}}$ satisfies the same equation as $F_{\bm{q}}$.
One can thus set $F_{-\bm{q}}=F_{\bm{q}}$ in Eq.~(\ref{eq:B9}).
Combining this equation with Eq.~(\ref{eq:B12}) finally yields Eqs.~(\ref{eq:Contact3}),
where the limit $\Lambda\to\infty$ is taken.

It is worthwhile to note that the same equation can be obtained in
a different way from a two-channel contact model, whose range parameter
is set to zero.

\subsubsection{Derivation of the mediated potentials}

The following provides the derivation of the Born-Oppenheimer potentials
Eq.~(\ref{eq:ContactSmallEAsymptote}) and Eq.~(\ref{eq:ContactLargeEAsymptote})
in the limit of small scattering length $(1/a\to-\infty)$ and unitarity
$(1/a\to0)$.

The Born-Oppenheimer equation (\ref{eq:ContactBornOppenheimer}) may
be written as follows,
\begin{equation}
E(r)=2n_{0}\left[\frac{2m}{4\pi\hbar^{2}}\frac{1}{a}+\frac{1}{V}\sum_{\bm{k}}\left(\frac{u_{k}^{2}}{E_{k}-E(r)}\left(1+\frac{\sin kr}{kr}\right)-\frac{1}{\epsilon_{k}}\right)\right]^{-1},\label{eq:D1}
\end{equation}

which can be further expressed as
\begin{equation}
E(r)=\frac{8\pi\hbar^{2}n_{0}}{2m}\left[\frac{1}{a}+\frac{2}{\pi}\int_{0}^{\infty}k^{2}dk\left(\frac{\frac{1}{2}\left(1+\frac{k^{2}+\xi^{-2}}{k\sqrt{k^{2}+2\xi^{-2}}}\right)}{k\sqrt{k^{2}+2\xi^{-2}}-\frac{2m}{\hbar^{2}}E(r)}\left(1+\frac{\sin kr}{kr}\right)-\frac{1}{k^{2}}\right)\right]^{-1}.\label{eq:D2}
\end{equation}

\paragraph*{Small scattering length}

In the limit of small scattering length $1/a\to-\infty$, the energy
$E$ goes to zero. Therefore, in this limit one can neglect the term
$E(r)$ in the denominator in the above equation. The integral can
then be calculated analytically, yielding the following explicit expression
for $E(r)$,
\begin{equation}
E(r)=\frac{8\pi\hbar^{2}n_{0}}{2m}\left[\frac{1}{a}-\frac{4+\pi}{2\sqrt{2}\pi}\frac{1}{\xi}+F(r)\right]^{-1}.\label{eq:D3}
\end{equation}
where $F(r)=[1+e^{-\sqrt{2}r/\xi}+2I_{0}(\sqrt{2}r/\xi)-2L_{0}(\sqrt{2}r/\xi)]/(4r)$,
and $I_{0}$ is the modified Bessel function of the first kind, and
$L_{0}$ denotes the modified Struve function. One can check that
$F(r)\approx\exp(-\frac{4+\pi}{2\sqrt{2}\pi}\frac{r}{\xi})/r$ for
$r\lesssim\xi$ and $F(r)\approx1/(4r)$ for $r\gg\xi$. However,
similary to the derivation of Appendix A, the convergence to Eq.~(\ref{eq:D3})
for small negative $a$ is not uniform. At very large distance, one
may not neglect the term $E(r)$ in the denominator, and one obtains
instead $F(r)=\frac{1}{r}\frac{2}{\pi}\int_{0}^{\infty}\tilde{F}(k)\sin krdk$
with $\tilde{F}(0)=\frac{\xi^{-1}}{2\sqrt{2}\vert\frac{2m}{\hbar^{2}}E(\infty)\vert}$,
giving the asymptotic behaviour,
\begin{equation}
F(r)\xrightarrow[r\to\infty]{}\frac{\xi^{-1}}{\pi\sqrt{2}\vert\frac{2m}{\hbar^{2}}E(\infty)\vert}\frac{1}{r^{2}},\label{eq:D4}
\end{equation}
which can be shown to hold in the present situation for $r\gg\xi^{-1}/\vert\frac{2m}{\hbar^{2}}E(\infty)\vert$.
As noted in Appendix A, because of the variational nature of the calculation,
the form of $E(r)$ at large distance is not physical, so we restrict
our consideration to $r\lesssim\xi$. Treating $a$ as a small perturbation,
one obtains
\begin{equation}
E(r)=\frac{8\pi\hbar^{2}n_{0}}{2m}\left[a+a^{2}\left(\frac{4+\pi}{2\sqrt{2}\pi}\frac{1}{\xi}-\frac{1}{r}\exp(-\frac{4+\pi}{2\sqrt{2}\pi}\frac{r}{\xi})\right)\right]+O(a^{3}),\label{eq:D5}
\end{equation}
which yields Eqs.~(\ref{eq:ContactSmallEAsymptote}). As noted in
the main text, this potential is close but somewhat different from
the perturbative result of Eq.~(\ref{eq:YukawaPotential}). If one
slightly worsens the variational potential by replacing $u_{k}$ by
$u_{k}-v_{k}$ in Eq.~(\ref{eq:D1}), one obtains a form that is
closer to Eq.~(\ref{eq:YukawaPotential}),
\begin{equation}
E(r)=\frac{8\pi\hbar^{2}n_{0}}{2m}\left[a+a^{2}\left(\sqrt{2}\frac{1}{\xi}-\frac{1}{r}\exp(-\sqrt{2}r/\xi)\right)\right]+O(a^{3}).\label{eq:D6}
\end{equation}

\paragraph*{Unitary limit}

For larger scattering lengths, the term $E$ becomes larger than $\frac{\hbar^{2}\xi^{-2}}{2m}$,
so that the integral in Eq.~(\ref{eq:D2}) is mostly determined by
$E(r)$ and we can treat $\xi^{-2}$ as a perturbation. To first order
in $\xi^{-2}$,we get:
\begin{equation}
\frac{\frac{1}{2}\left(1+\frac{k^{2}+\xi^{-2}}{k\sqrt{k^{2}+2\xi^{-2}}}\right)}{k\sqrt{k^{2}+2\xi^{-2}}-\frac{2m}{\hbar^{2}}E}=\frac{1}{k^{2}+\kappa^{2}}+\frac{1}{(k^{2}+\kappa^{2})^{2}}\xi^{-2}+O(\xi^{-4})\label{eq:D7}
\end{equation}
where we set $E=-\hbar^{2}\kappa^{2}/(2m)$. The integration can then
be performed, yielding
\begin{equation}
E(r)=\frac{8\pi\hbar^{2}n_{0}}{2m}\left[\frac{1}{a}+\left[-\kappa+\frac{\exp(-\kappa r)}{r}+\left(-\frac{1+\exp(-\kappa r)}{2\kappa}\right)\xi^{-2}\right]\right]^{-1}.\label{eq:D8}
\end{equation}
or equivalently,
\begin{equation}
\frac{1}{a}-\kappa-\frac{\xi^{-2}}{2\kappa}+\left(\frac{1}{r}-\frac{\xi^{-2}}{2\kappa}\right)\exp(-\kappa r)=-\frac{8\pi n_{0}}{\kappa^{2}}.\label{eq:D9}
\end{equation}
from which Eq.~(\ref{eq:ContactLargeE}) is obtained. 

Now we take the unitary limit $1/a\to0$ in Eq.~(\ref{eq:D9}). For
small $r$, the term $1/r$ dominates over the other terms except
$\kappa$ so that the equation reduces to
\begin{equation}
-\kappa r+e^{-\kappa r}=0.\label{eq:D10}
\end{equation}
It can be checked that the missing terms can be neglected when $r\ll L$
(and $r\ll\xi$ which is readily satisfied since $L\ll\xi$). The
solution of Eq.~(\ref{eq:D10}) is $\kappa r=W(1)$, where $W$ is
the Lambert function, leading to the result of Eq.~(\ref{eq:ContactLargeEAsymptote}a).
For large $r$, one can first neglect the $r$-dependent terms since
they vanish. One obtains the threshold value $\kappa_{\infty}$ satisfying
the equation
\begin{equation}
\kappa_{\infty}+\frac{\xi^{-2}}{2\kappa_{\infty}}=\frac{L^{-3}}{\kappa_{\infty}^{2}}\label{eq:D11}
\end{equation}
This cubic equation in $\kappa_{\infty}$ admits the solution
\begin{equation}
\kappa_{\infty}=X^{1/3}-\frac{\xi^{-2}}{6X^{1/3}}\label{eq:D12}
\end{equation}
with $X=\frac{1}{2}\left(L^{-3}+\sqrt{L^{-6}+\frac{4}{27}(\frac{1}{2}\xi^{-2})^{3}}\right)=L^{-3}+O(\xi^{-6})$.
Treating the $r$-dependent terms of Eq.~(\ref{eq:D9}) as a perturbation,
one can set $\kappa(r)=\kappa_{\infty}(1+\varepsilon(r))$ in this
equations, which yields $\varepsilon(r)=\left(3\kappa_{\infty}+\frac{\xi^{-2}}{2\kappa_{\infty}}\right)^{-1}\left(\frac{1}{r}-\frac{\xi^{-2}}{2\kappa_{\infty}}\right)e^{-\kappa_{\infty}r}$.
This gives, to first order in $\xi^{-2},$ the result of Eq.~(\ref{eq:ContactLargeEAsymptote}b).
The smallness of $\varepsilon(r)\ll1$ requires that $r\gg\kappa_{\infty}^{-1}\sim L$.

Again, we should note that according to Eq.~(\ref{eq:D4}) one finds
that for much larger distances (in this case for $r\gg\xi$) the potential
approaches its threshold as $1/r^{2}$, although this is an unphysical
artifact of the variational ansatz. 

\subsubsection{Numerical solution}

In the $s$-wave channel ($F_{\bm{q}}=F_{q}$), where the Efimov attraction
takes place, the equation (\ref{eq:Contact3}) simplifies as follows,
\begin{equation}
\left(\frac{1}{T_{q}(E)}+\frac{2n_{0}}{2\varepsilon_{q}-E}\right)F_{q}+\frac{1}{V}\sum_{\bm{k}}^{k<\Lambda_{3}}\frac{u_{\vert\bm{k}+\bm{q}\vert}^{2}\;}{E_{\vert\bm{k}+\bm{q}\vert}+\varepsilon_{k}+\varepsilon_{q}-E}F_{k}=0,\label{eq:B14}
\end{equation}
where a three-body momentum cutoff $\Lambda_{3}$ has been imposed
on the argument of $F$. Making the substitution $\frac{1}{V}\sum_{\bm{k}}\approx(2\pi)^{-3}\int d^{3}\bm{k}$
and using the explicit forms of $T_{k}(E)$, $E_{k}$ and $\varepsilon_{k}$
gives 
\begin{equation}
M_{q}(z)F_{q}+\int_{0}^{\Lambda_{3}}dkM_{qk}(z)F_{k}=0.\label{eq:B15}
\end{equation}
where
\begin{equation}
M_{q}(z)=\frac{1}{a}+\frac{2}{\pi}\int_{0}^{\infty}dk\left(\frac{M}{2\mu}\frac{ku_{k}^{2}}{q}\text{arctanh}\frac{2kq}{\frac{M}{m}\sqrt{k^{2}(k^{2}+2\xi^{-2})}+\left(k^{2}+2q^{2}\right)-\frac{M}{\mu}z}-1\right)+\frac{8\pi n_{0}}{\frac{2\mu}{M}q^{2}-z},\label{eq:B16}
\end{equation}
\begin{equation}
M_{qk}(z)=\frac{1}{\pi}\frac{k}{q}\int_{\vert k-q\vert}^{\vert k+q\vert}\frac{pu_{p}^{2}}{\frac{\mu}{m}p\sqrt{p^{2}+2\xi^{-2}}+\frac{\mu}{M}(k^{2}+q^{2})-z}dp,\label{eq:B17}
\end{equation}
and $z=\frac{2\mu}{\hbar^{2}}E$. Equation (\ref{eq:B15}) can be
solved as a matrix problem by discretising the momenta $q$ and $k$
on a a grid. The eigenvalues can be found by standard linear algebra
routines, and the energy levels are obtained by finding the values
of $z$ which make one of the eigenvalues equal to zero.

 \end{widetext}
\end{document}